\documentclass[12pt,preprint]{aastex}

\newcommand{\lta}{{\>\rlap{\raise2pt\hbox{$<$}}\lower3pt\hbox{$\sim$}\>}}
\newcommand{\gta}{{\>\rlap{\raise2pt\hbox{$>$}}\lower3pt\hbox{$\sim$}\>}}

\begin{document}

\title{RELATIONS BETWEEN THE LUMINOSITY, MASS, AND AGE DISTRIBUTIONS
       OF YOUNG STAR CLUSTERS}

\author{S. Michael Fall}
\affil{Space Telescope Science Institute, 3700 San Martin Drive, 
Baltimore, MD 21218; fall@stsci.edu}

\begin{abstract}

We derive and interpret some relations between the luminosity,
mass, and age distributions of star clusters, denoted here
by $\phi(L)$, $\psi(M)$, and $\chi(\tau)$, respectively.
Of these, $\phi(L)$ is the easiest to determine observationally,
whereas $\psi(M)$ and $\chi(\tau)$ are more informative about
formation and disruption processes.
For a population of young clusters, with a relatively wide
range of ages, $\phi(L)$ depends on both $\psi(M)$ and 
$\chi(\tau)$ and thus cannot serve as a proxy for $\psi(M)$
in general.
We demonstrate this explicitly by four illustrative examples 
with specific forms for either $\psi(M)$ or $\chi(\tau)$.
In the special case in which $\psi(M)$ is a power law and is 
independent of $\chi(\tau)$, however, $\phi(L)$ is also 
a power law with the same exponent as $\psi(M)$.
We conclude that this accounts for the observed similarity 
between $\phi(L)$ and $\psi(M)$ for the young clusters in the 
Antennae galaxies.
This result reinforces our picture in which clusters form with 
$\psi(M) \propto M^{-2}$ and are then disrupted rapidly at a 
rate roughly independent of their masses.
The most likely disruptive process in this first stage is the
removal of interstellar matter by the energy and momentum input 
from young stars (by photoionization, winds, jets, and 
supernovae).
The few clusters that avoid this ``infant mortality'' are
eventually disrupted in a second stage by the evaporation of 
stars driven by two-body relaxation, a process with a strong
dependence on mass.
We suspect this picture may apply to many, if not all,
populations of star clusters, but this needs to be verified 
observationally by determinations of $\psi(M)$ and 
$\chi(\tau)$ in more galaxies. 

\end{abstract}



\section{INTRODUCTION}

In this paper, we derive and interpret some relations
between the luminosity, mass, and age distributions of
young star clusters.
We define these ``distributions'' (probability densities, 
in fact) as follows:
$\phi(L)dL$ is the fraction of clusters with luminosities 
between $L$ and $L + dL$, 
$\psi(M)dM$ is the fraction of clusters with masses between
$M$ and $M + dM$, and
$\chi(\tau)d\tau$ is the fraction of clusters with ages between
$\tau$ and $\tau + d\tau$.
We focus here on the question: To what extent does $\phi(L)$
reflect $\psi(M)$, and how does this depend on $\chi(\tau)$?

For a population of clusters with the same mass-to-light
ratio, resulting from a common age and stellar mass function
(and neglecting any mass dependence of stellar escape rates), 
$\phi(L)$ always has the same form as $\psi(M)$. 
This assumption is often made in studies of old globular 
clusters, when their colors or other evidence indicates
that the spread in ages is modest or narrow (in a fractional 
sense).
However, for a population of clusters with a wide range of
mass-to-light ratios, $\phi(L)$ can be very different from
$\psi(M)$.
This is the generic case when the spread in ages is broad in 
fractional terms, as happens when clusters form continually
up to the present time. 
We are concerned in this paper primarily with this situation.

The luminosity function of a population of clusters is 
relatively easy to determine because it requires observations 
in only one photometric band. 
Consequently, there is now an extensive literature on this 
subject.
In studies with deep observations of large samples of young
clusters, the luminosity function is often found to have an
approximate power-law form, $\phi(L) \propto L^{\alpha}$, with 
an exponent near $\alpha \approx -2$.
Early examples of this result include the Milky Way (van den
Bergh \& LaFontaine 1984), the Large Magellanic Cloud (Elson 
\& Fall 1985), and Messier 33 (Christian \& Schommer 1988).
A more recent example, and the one that motivates the work
presented here, is the interacting Antennae galaxies (Whitmore 
et al. 1999).
In some cases, the luminosity function is a power law in a 
first approximation, but also appears to have some secondary
convex curvature in log-log plots.
However, this curvature often has only marginal statistical 
significance, and conceivably it could be the result of subtle 
systematic effects, such as undercorrections for incompleteness 
or photometric errors at faint magnitudes.
We ignore such curvature here in the interest of keeping our 
analysis as simple and transparent as possible.   

Because the luminosity function is affected by the fading of 
the stellar populations within the clusters, it does not tell
us directly about the formation and disruption of the clusters. 
The mass and age distributions are more fundamental in this 
regard.
In particular, for the youngest clusters, $\psi(M)$ is likely
to reflect fairly directly the physical processes involved in
the formation of the clusters and/or their parent molecular 
clouds.
The initial mass function of clusters thus plays a role in 
the theory of cluster formation similar to that of the initial
mass function of stars in the theory of star formation.
The age distribution, in principle, reflects a combination 
of the formation and disruption histories of the clusters.
In practice, however, $\chi(\tau)$ is primarily a diagnostic 
of disruption processes, because it usually has a stronger
dependence on $\tau$ than is plausible for the formation
history. 

The mass and age distributions require photometry in 
several bands, corrections for interstellar redding,
and comparisons with stellar population models.
This makes their determination more laborious than that
of the luminosity function.
Moreover, estimating two univariate distributions or one 
bivariate distribution (of $M$ and $\tau$) to a given level 
of statistical accuracy requires a larger sample of clusters 
than estimating a single univariate distribution (of $L$) to
the same accuracy.
Consequently, we have reliable determinations of the mass and
age distributions of clusters in only a few galaxies, the best 
example again being the Antennae.
In this case, the mass and age distributions can be 
represented by the power laws $\psi(M) \propto M^{\beta}$ 
with $\beta \approx -2$ (Zhang \& Fall 1999) and $\chi(\tau) 
\propto \tau^{\gamma}$ with $\gamma \approx -1$ (Fall,
Chandar, \& Whitmore 2005).
The exponent of $\chi(\tau)$ quoted here is for mass-limited
samples, as is required in all the formulae of this paper;
for luminosity-limited samples, $\chi(\tau)$ is significantly
steeper, as a consequence of the rapid fading of the clusters.
Furthermore, over the observed domain of masses and ages,
$\psi(M)$ and $\chi(\tau)$ are approximately independent of 
each other (as shown in the previous references).
This fact simplifies much of the analysis in this paper.

The observations of the Antennae galaxies raise an 
interesting question: Why do the luminosity and mass 
functions of the young clusters have the same, or at least 
approximately the same form? i.e., why are they both 
power laws with the same exponent?
In much of the literature on star clusters, there is a 
tendency to regard the luminosity function as a proxy for 
the mass function.
As noted above, this cannot be true in general, given that 
clusters of different ages have different mass-to-light ratios. 
Yet the evidence from the Antennae galaxies suggests that 
$\phi(L)$ and $\psi(M)$ are in fact closely related.
Why is this? Answering this question is the main purpose
of this paper. 
Although our analysis is motivated by observations of the
Antennae galaxies, it seems likely on theoretical grounds
that our conclusions have wider applicability, possibly to
most or even all galaxies with large populations of young 
clusters.

\section{GENERAL RELATIONS}

The general relation between the luminosity and mass functions 
of star clusters must also involve the age distribution.
With this in mind, we introduce the following bivariate 
distributions: $f(L, \tau)dLd\tau$ is the fraction of clusters 
with luminosities between $L$ and $L + dL$ and ages between 
$\tau$ and $\tau + d\tau$, and $g(M, \tau)dMd\tau$ is the 
fraction of clusters with masses between $M$ and $M + dM$ 
and ages between $\tau$ and $\tau + d\tau$.
In terms of these distributions, the univariate 
distributions of luminosity, mass, and age are 
$$
\phi(L) = \int_0^{\infty} f(L, \tau) d\tau,
\eqno(1)
$$
$$
\psi(M) = \int_0^{\infty} g(M, \tau) d\tau,
\eqno(2)
$$
$$
\chi(\tau) = \int_0^{\infty} f(L, \tau) dL
           = \int_0^{\infty} g(M, \tau) dM.
\eqno(3)
$$

We now make our {\it first simplifying assumption, that the 
mass-to-light ratios of clusters, denoted by $\mu$, depend 
only on their ages:}
$$
\mu(\tau) = M(\tau)/L(M, \tau).
\eqno(4)
$$ 
In principle, $\mu$ could vary among clusters of the same
age if, for example, they had different stellar mass functions.
In practice, such variations are found or assumed to be small,
and equation~(4) is thus the basis for all determinations of 
the mass functions of clusters from multiband photometry.
In this case, the bivariate distributions are related by
$$
f(L, \tau) = g(M, \tau) |(\partial{M}/\partial{L})_{\tau}|
           = g(M, \tau) \mu(\tau).
\eqno(5)
$$
Combining equations (1), (4), and (5) yields
$$
\phi(L) = \int_0^{\infty} g[\mu(\tau)L, \tau] \mu(\tau) d\tau.
\eqno(6)
$$
Given $g(M, \tau)$, equations (2), (3), and (6) determine fully
the relations between $\phi(L)$, $\psi(M)$, and $\chi(\tau)$.
These relations in general are quite complicated.

We now make our {\it second simplifying assumption, that the 
mass and age distributions of clusters are independent of each 
other:}
$$
g(M, \tau) = \psi(M) \chi(\tau).
\eqno(7)
$$
This is valid if the rates of formation and disruption of the
clusters are independent of their masses. 
For very young clusters, which dominate the bright parts of
$\phi(L)$, the main disruptive effect is loss of interstellar 
matter by the energy and momentum input from young stars 
(photoionization, stellar winds and jets, and supernovae 
remnants). 
These inputs should be roughly proportional to the masses of the 
clusters and hence also to the amount of material to be removed.
Thus, we expect the disruption rate to be roughly independent of 
the masses of the clusters, as prescribed by equation~(7) above 
(see Fall et al. 2005 for further discussion of this and related 
issues). 
This is consistent with our observations of the young clusters 
in the Antennae galaxies, where we have determined $g(M, \tau)$
for $\tau \la 10^7 (M/10^4 M_{\odot})^{1.3}$~yr.
Once the evaporation of stars by two-body relaxation 
becomes important, the disruption rate is mass-dependent, and 
equation~(7) is no longer valid.
This occurs for $\tau \ga 5 \times 10^8 (M/10^4 M_{\odot})$~yr,
i.e., well outside the observed mass-age domain of clusters in 
the Antennae galaxies (see Section~4 below).

When we insert equation (7) into equation (6), we obtain
$$
\phi(L) = \int_0^{\infty} \psi[\mu(\tau)L] \chi(\tau) 
          \mu(\tau) d\tau.
\eqno(8)
$$
It is helpful at this point to use the mass-to-light ratio 
$\mu$ as a substitute for the age $\tau$. With this in mind, 
we introduce the corresponding distribution, defined such 
that $\theta(\mu) d\mu$ is the fraction of clusters with 
mass-to-light ratios between $\mu$ and $\mu + d\mu$. 
This is related to the age distribution by
$$
\theta(\mu) = \chi(\tau) |d\tau/d\mu|.
\eqno(9)
$$
Combining this with equation~(8) gives
$$
\phi(L) = \int_0^{\infty} \psi(\mu L) \theta(\mu) \mu d\mu.
\eqno(10)
$$
This is the simplest form of the general relation between the 
luminosity, mass, and age (or mass-to-light ratio) distributions 
subject to the two simplifying assumptions embodied in equations
(4) and~(7). 
To make further progress, we must specify $\psi(M)$ and/or
$\theta(\mu)$.

\section{ILLUSTRATIVE EXAMPLES}

We now consider four simple, illustrative examples, based on 
specific assumptions about the distributions of masses and
mass-to-light ratios (delta functions and power laws). 
We could extend the list of examples indefinitely, but those
presented here are sufficient to illustrate our main conclusions.
 
{\it Example 1: delta-function distribution of mass-to-light 
ratios, $\theta(\mu) = \delta(\mu - \mu_0)$, and arbitrary 
distribution of masses, $\psi(M)$.}
In this case, all clusters have the same mass-to-light ratio 
$\mu_0$ and hence the same age $\tau_0$  given by $\mu_0 = 
\mu(\tau_0)$.
The luminosity function, from equation (10), is 
$$
\phi(L) = \mu_0 \psi(\mu_0 L).
\eqno(11)
$$
Here, $\phi(L)$ and $\psi(M)$ are identical apart from a 
rescaling of their arguments by $\mu_0$. 
This is appropriate for populations of clusters that are
observed long after they formed, with typical ages much larger 
than their age spreads (i.e., $\tau_0 \gg \Delta\tau$).
This is probably a reasonable approximation for old globular
clusters. 
However, it is a poor approximation for populations of clusters 
that are still forming, such as that in the Antennae galaxies, 
since in this case the typical age is comparable to the age 
spread (i.e., $\tau_0 \sim \Delta\tau$).

{\it Example 2: power-law distribution of mass-to-light ratios, 
$\theta(\mu) = A \mu^{\epsilon}$, and arbitrary distribution 
of masses, $\psi(M)$.} 
According to equation~(10), the luminosity function is
$$
\phi(L) = A L^{-(2 + \epsilon)} \int_0^{\infty} \psi(M) 
          M^{1 + \epsilon} dM.
\eqno(12)
$$
In this case, $\phi(L)$ is a power law with exponent $\alpha
= - (2 + \epsilon)$ irrespective of the form of $\psi(M)$.
This example illustrates in an extreme way that $\phi(L)$, in 
general, depends on $\chi(\tau)$, and not on $\psi(M)$ alone.
Equation~(12) is valid so long as the integral on the right-hand 
side exists. 
This condition is satisfied whenever $\psi(M)$ is finite over 
a finite range of $M$ and zero elsewhere, as is true for real 
populations of star clusters.
A power-law model of $\psi(M)$, if extended both to $M = 0$
and $M \rightarrow \infty$, would, however, cause an artificial
divergence of the integral in equation~(12).

It is worth exploring this example in a little more detail.
Stellar population models indicate that the mass-to-light 
ratio evolves approximately as a power law, $\mu(\tau) \propto 
\tau^{\delta}$, over a wide range of ages, $\tau \ga {\rm few} 
\times 10^6$~yr (after smoothing over many small wiggles), with 
$\delta \approx 0.7$ for luminosities in the $V$ band, and larger
(smaller) exponents for shorter (longer) wavelengths (Leitherer 
et al. 1999; Bruzual \& Charlot 2003; see e.g. Fig.~1 of Fall 
et al. 2005). 
Assuming that the age distribution is also a power law,
$\chi(\tau) \propto \tau^{\gamma}$, we then have, from 
equations~(9) and~(12), $\theta(\mu) \propto \mu^{\epsilon}$ 
with $\epsilon = (\gamma - \delta + 1)/\delta$, and $\phi(L) 
\propto L^{\alpha}$ with 
$$
\alpha = - 1 - (1 + \gamma)/\delta.
\eqno(13)
$$
In this case, the exponent of $\phi(L)$ depends on the
exponent of $\chi(\tau)$ and, except in the special case 
$\gamma = -1$, on the photometric band (through $\delta$).

For the young clusters in the Antennae galaxies, equation~(13) 
and the observed age distribution, with $\gamma \approx -1$, 
predict $\alpha \approx -1$, and hence a luminosity function
significantly shallower than the observed one, with $\alpha
\approx -2$.
Equation~(12), therefore, does {\it not} explain the observed 
form of $\phi(L)$ in the Antennae galaxies.
We emphasize that this conclusion depends crucially on the
observed {\it decline} of $\chi(\tau)$.
In the absence of this information, equation~(12) {\it would} 
provide an acceptable explanation for the observed form 
of $\phi(L)$, as the following example shows.
For $\gamma \approx 0$, corresponding to $\chi(\tau)
\approx {\rm const}$ and hence little if any disruption of 
clusters, equation~(13) implies $\alpha \approx -2$, close 
to the observed exponent of $\phi(L)$ in many galaxies. 
Thus, without evidence to the contrary, we could speculate
that nearly uniform age distributions, rather than power-law 
mass functions, are responsible for the observed power-law 
luminosity functions.\footnote{
See Hogg \& Phinney (1997) for further discussion of this 
possibility in the context of the luminosity function of 
galaxies.}

{\it Example 3: delta-function distribution of masses, 
$\psi(M) = \delta(M - M_0)$, and arbitrary distribution of 
mass-to-light ratios, $\theta(\mu)$.}
In this case, all clusters have the same mass $M_0$, and the
luminosity function, from equation~(10), is
$$
\phi(L) = M_0 L^{-2} \theta(M_0/L).
\eqno(14)
$$
Here, as in the previous example, $\phi(L)$ is determined 
entirely by $\theta(\mu)$ and hence by $\chi(\tau)$. 
Equation~(14) also demonstrates that $\phi(L)$ can be 
a broad function even when $\psi(M)$ has no width at all.
In particular, if $\theta(\mu)$ is a power law with exponent 
$\epsilon$, then $\phi(L)$ is a power law with exponent 
$\alpha = -(2 + \epsilon)$, as expected, because equation~(14) 
is then a special case of equation~(12).
Once again, this is not a viable explanation for the
observed luminosity function of young clusters in the 
Antennae galaxies.

{\it Example 4: power-law distribution of masses, $\psi(M) = 
B M^{\beta}$, and arbitrary distribution of mass-to-light 
ratios, $\theta(\mu)$.} 
According to equation~(10), the luminosity function is
$$
\phi(L) = B L^{\beta} \int_0^{\infty} \theta(\mu) 
          \mu^{1 + \beta} d\mu.
\eqno(15)
$$
In this case, $\phi(L)$ and $\psi(M)$ have identical forms; 
both are power laws with the same exponent, irrespective of 
$\chi(\tau)$ and the band in which luminosities are
measured.
This probably {\it is} a good description of the population 
of young clusters in the Antennae galaxies, for which the 
observed luminosity, mass, and age distributions are 
approximate power laws with exponents $\alpha \approx 
\beta \approx -2$ and $\gamma \approx -1$, respectively.
In this case, as with any power-law model of $\chi(\tau)$,
one might wonder whether the integral in equation~(15) 
converges at small $\mu$, corresponding to small $\tau$.
However, in all the photometric bands of interest here
(near UV through near IR), $\mu(\tau)$ follows a power 
law down to a finite, minimum value $\mu_{\rm min}$, 
which occurs at $\tau \approx {\rm few} \times 10^6$~yr 
(Leitherer et al. 1999; Bruzual \& Charlot 2003).
This truncates $\theta(\mu)$ for $\mu < \mu_{\rm min}$
and ensures that the integral is finite.

\section{DISCUSSION AND CONCLUSIONS}

We can now answer the question posed in the Introduction: 
Why does the luminosity function of young clusters in the 
Antennae galaxies have the same, or approximately the same
form as the mass function?
As the examples in the previous section show, the answer has 
two parts: (1) because the mass function is a power law, and
(2) because it is statistically independent of the age 
distribution.
The exact form of $\chi(\tau)$ is immaterial in the relation
between $\phi(L)$ and $\psi(M)$ provided only that it declines
steeply enough that Example~2 above is irrelevant.
The independence of $\psi(M)$ and $\chi(\tau)$, however, is 
a key element.
This may at first seem like only a mathematical convenience, 
but in fact it has important physical implications, as the 
following discussion makes clear.

Star clusters are disrupted by a variety of processes,
beginning with the removal of interstellar matter by stellar 
activity (photoionization, winds, jets, supernovae), leading
to ``infant mortality.''
The later disruptive effects include stellar mass loss,
dynamical friction, gravitational shocks, and stellar evaporation
driven by two-body relaxation (see Spitzer 1987 for a review). 
In massive galaxies like the Milky Way and the Antennae, dynamical 
friction and gravitational shocks, while potentially important 
for some clusters near the centers of the galaxies, have relatively 
little effect on most of the other clusters.
Moreover, the mass loss by stellar evolution and gravitational 
shocks, even when these processes are important, tends to preserve 
the shape of the mass function of the clusters; in particular, 
an initial power law remains a power law with the same exponent 
[see eqs.~(12) and (13) of Fall \& Zhang 2001].

The dominant long-term disruptive effect for most clusters 
is the gradual escape of stars resulting from gravitational 
scattering by other stars within the same clusters 
(``evaporation'' driven by two-body relaxation).
We denote the rate of mass loss $-dM/dt$ by this process by
$\mu_{\rm ev}$ (not to be confused with a mass-to-light ratio).
For clusters with a constant mean internal density, set by the
smooth tidal field of their host galaxy, $\mu_{\rm ev}$ also
remains constant, and $M$ decreases linearly with time $t$. 
If the clusters of current age $\tau$ formed at a rate 
$h(\tau)$ in the past, with an initial mass function 
$\psi_0(M)$, then the current bivariate distribution of 
masses and ages is given by 
$$
g(M, \tau) = \psi_0(M + \mu_{\rm ev}\tau) h(\tau).
\eqno(16)
$$
[This is a straightforward generalization of eq.~(11) of
Fall \& Zhang 2001.]
Evidently, $g(M, \tau)$ is independent of $M$ for $M \ll
\mu_{\rm ev}\tau$, has a turnover at $M \approx \mu_{\rm ev}\tau$,
and is proportional to $\psi_0(M)$ for $M \gg \mu_{\rm ev}\tau$.
This form of $g(M, \tau)$ is not covered by the examples in the 
previous section because it is not the product of a function 
of $M$ alone and a function of $\tau$ alone, as in equation~(7).
We note incidentally that this is true of {\it any} disruptive 
process for which the mass-loss rate of a cluster depends on 
its mass, as in all models of the form $dM/dt = - M/t_d(M)$ 
with $t_d(M) \propto M^k$, except the special case $k = 0$.

It is now interesting to consider the clusters in the 
Antennae galaxies in the context of equation~(16). 
Detailed dynamical calculations for massive galaxies like
the Milky Way and the Antennae give $\mu_{\rm ev} \approx 
2 \times 10^{-5} M_{\odot} {\rm yr}^{-1}$ and hence a
turnover or ``peak'' in the mass function of the clusters 
at $M_p \approx 2 \times 10^5 (\tau/10^{10}{\rm yr}) 
M_{\odot}$ (Fall \& Zhang 2001).
This means that the clusters most affected by stellar 
evaporation have $\tau \ga M/\mu_{\rm ev} \approx 5 
\times 10^8 (M/10^4 M_{\odot})$~yr and thus are not 
included in the empirical determination of $g(M, \tau)$.
We are thus led to a two-stage picture for the disruption
of the clusters (Fall et al. 2005).
In the first stage, the energy and momentum input from 
massive young stars removes the interstellar matter from 
protoclusters, leaving many of them gravitationally unbound 
(``infant mortality'').
The empirical evidence is that this process disrupts most
of the clusters relatively rapidly but largely independently
of their masses, thus reducing the amplitude of the mass
function while preserving its initial power-law shape.
Since nearly all the observed clusters are in this phase
of evolution, the illustrative examples of the previous 
section apply, as does our explanation for the similarity 
between the luminosity and mass functions. 
The few clusters that survive the first stage are then 
liable to disruption in the second by stellar evaporation 
driven by two-body relaxation.
As noted above, this process eventually changes the shape 
of the mass function into that of old globular clusters.

How general is this picture?
Does it apply to the clusters in all galaxies, or only 
to those in a small subset of galaxies like the Antennae?
The initial mass function of star clusters is probably
similar to the mass spectrum of their parent molecular 
clouds, and it is possible that this is determined by  
nearly universal, fractal-like density and/or velocity 
fields in the interstellar medium (Elmegreen 2002 and 
references therein).
Moreover, the disruption rates in both the first (rapid) 
and second (slow) stages discussed above depend mainly on 
the properties of the clusters, not those of their host 
galaxies.
Thus, we expect the picture outlined here to be fairly
general.
Nevertheless, it is important to test it observationally
in detail in at least a few more cases, especially in 
non-interacting, quiescent galaxies.
This requires determinations of the mass and age
distributions over wide ranges of mass and age, which 
in turn requires deep multiband photometry of large 
samples of clusters.
This should be a higher priority than simply
determining the luminosity functions of clusters
in more galaxies. 

Our main conclusion, at least for the young 
clusters in the Antennae galaxies, is that the 
luminosity function has the same form as the mass 
function because the latter is a power law and is 
independent of the age distribution.
But the main lesson of this paper is that 
disentangling the relations between the luminosity, 
mass, and age distributions has helped to clarify 
our picture of the formation and disruption of the 
clusters.

The author is grateful to Rupali Chandar, Michael Santos,
Bradley Whitmore, and the referee for helpful comments.

\end{document}